\documentstyle[preprint,aps]{revtex}

\begin{document}
\draft
\preprint{IMSc/2002/05/09}
\title{Quantum Geometric Description of Cosmological Models}
\author{G. Date \footnote{e-mail: shyam@imsc.ernet.in}}
\address{The Institute of Mathematical Sciences,
CIT Campus, Chennai-600 113, INDIA.}
\maketitle

\begin{abstract}
This is a written version of the review talk given at the meeting on 
``Interface of Gravitational and Quantum Realms" at IUCAA, Pune during 
December 2001. The talk reviewed the recent work of Martin Bojowald on Loop 
Quantum Cosmology.
\end{abstract}

\vskip 0.5cm


\narrowtext			

\section{Introduction}		

Canonical quantization of gravity formulated in terms of connection
variable is now at a stage where one can address questions of physical
interest. At the interpretational level, one still has the outstanding
issues of understanding the classical limit and development of a
systematic semiclassical expansion.\cite{thiemann} \\

There are at least two classes of situations in classical GR that call for 
a quantum elucidation: (i) situations where classical GR predicts 
`singularities' eg early universe and collapse, and (ii) situations where 
GR predicts horizons possessing entropy. In a sense, these can be thought of 
as `requiring' a quantum theory of gravity and hence any such theory should 
lead to better understanding of these situations. \\

There is another context in which one can look for signatures of quantum
nature of space-time. This refers to matter wave propagation in a
quantum geometrical background. Looking for such effects experimentally
actually seems possible thanks to the GRB sources at cosmological
distances. This context is very different from the first two in that QG
effects are being probed in very ordinary, highly classical, non-extreme 
situations.\cite{urrutia}  \\

The horizon context has already been analyzed and quantum geometry
framework does provide a microscopic understanding of entropy. Hawking
effect is however is not yet clear.\cite{entropy}  \\

Of the two singular context, Martin Bojowald has recently analyzed the
cosmological context in detail. As this is a first application of the
quantum geometry framework with very interesting results, I decided to
review this work in some details.

\section{Standard Quantum Cosmology}

A version of quantum gravity in the cosmological context was already
attempted in the sixties. This was again a canonical quantization approach
but using the geometro-dynamical variables - metric $g_{ij}$ and a symmetric 
second rank tensor $k_{ij}$ defined on a three manifold. This is a
constrained system with diffeomorphism constraint implementing spatial
diffeomorphism and the Hamiltonian constraint implementing the
space-time diffeomorphisms. Since one does not know how to carry out the Dirac 
(or otherwise ) quantization of this constrained system, one looks for
highly symmetric class of space-times, does the symmetry reduction
classically and quantizes the left over finitely many degrees of
freedom. In the context of so called Bianchi models, the gravitational 
phase space is at most six dimensional. This is a `minisuperspace'
quantization and has become synonymous with `standard quantum
cosmology'.\cite{std-cosmology} Here are its main conclusions.\\

(i) The Hamiltonian constraint (quadratic in momenta) can be interpreted
as an evolution equation which is second order in a suitably chosen `time' 
variable. This is intimately connected with the `problem of time'. In a
4-diffeomorphism invariant theory one has to interpret dynamics in
terms of evolution relative to a `clock' degree of freedom. For FRW
models, the scale factor (a gravitational degree of freedom) is a
natural choice of a `clock'. The evolution equation - Wheeler-De Witt
equation - being second order in time has two independent solutions i.e.
a non-unique wave function of the universe. \\

(ii) When the scale factor vanishes, the inverse scale factor and hence the 
curvatures blow up, implying persistence of classical singularity. \\

(iii) All geometrical quantities such as areas, volumes etc have
continuous spectra. \\

Note that what is quantized is a {\it finite} dimensional phase space as
nothing else could be done. One was forced to do the symmetry reduction
classically and then proceed to quantization. The Ashtekar reformulation in
terms of connection variables offers a different alternative. One can
quantize the infinite dimensional kinematical phase space and explore the 
possibility of doing symmetry reduction after quantization.  Further more
generically the spectra of areas , volumes etc turn out to be discrete
indicating that this quantization is qualitatively different. 

\section{Quantum Geometry Framework}

Let me briefly recall some of the basic steps. \\

(a) Choose a classical phase space: This involves choosing a three
manifold, $\Sigma$, with or without boundary and/or asymptotic regions.
The basic variables (for the gravitational sector) are $su(2)$ (Lie algebra 
of $SU(2)$ group) valued
connection 1-forms, $A_{a}^{i}\tau_{i}$ and a densitised triad,
$E_{i}^{a}$ (`dual' to a 2-form). The classical configuration space is
the space of all smooth (real analytic) connections satisfying appropriate
boundary conditions, $\cal{A}$. The space modulo gauge transformations
is denoted as $\cal{A/G}$. The phase space is the cotangent bundle of
the configuration space. On this phase space we have the usual Gauss
law, the diffeomorphism and the Hamiltonian constraints.\\

(b) There is a natural class of functions on $\cal{A/G}$ namely, $Tr(P
exp \int_{\gamma} A )$. These are labeled by closed loops in $\Sigma$,
are gauge invariant and diffeomorphism invariant. One now chooses to
look for a Hilbert space on which these functions are represented by
multiplicative operators. This is achieved by using the commutative
$C^*$ algebra  of these functions and choosing one of the
representations of this algebra. This gives $H_{kin} = 
L^{2}(\overline{\cal{A/G}},
\mu_{AL})$. In other words, the Hilbert space is the space of square
integrable complex functions on the `quantum configuration space'
$\bar{\cal{A/G}}$ square integrable with respect to the Ashtekar-Lewandowski 
measure constructed from the Haar measure on $SU(2)$. A convenient
description of the Hilbert space is obtained in terms of the so called
spin-network functions.\cite{AL} \\

In $\Sigma$ consider closed graphs $\gamma$. Associate with each edge $e_i$
a representation $\pi_i$ of $SU(2)$ and associate with each vertex $v_{\alpha}$
an intertwiner (contractor/invariant tensor), $C_{\alpha}$. For each $A \in
\bar{\cal{A}}$ one has, by definition, $g_i = A(e_i)$, the generalized
holonomy. Define,
\begin{equation}
\psi_{\gamma}(A) = ``Tr" \Pi_{edges} \pi_i(g_i) \ \Pi_{vertices} C_{\alpha}
\end{equation}

These are the spin network functions - functions on $\cal{A/G}$ -
labeled by graphs, representations and contractors. The set of all such
functions forms an orthonormal set which is dense in $H_{kin}$. In
practice one defines various operators by their actions on these and
extends them to $H_{kin}^{\Sigma}$. This dense space is also called the space of
cylindrical function.\\

(c) In order to accommodate the possibility that the zero eigenvalue of
constraints could be in a continuum (i.e. a generalized eigenvalue) one
introduces a `rigging' - $\Omega \subset H_{kin} \subset \Omega^*$. $\Omega$
is the dense space above while $\Omega^*$ is the space of continuous
linear functional on $\Omega$. The important point is that the physical
states generically belong to $\Omega^*$. One says that physical states
are distributional.\cite{diff-inv} \\

(d) For matter sector analogous constructions based on suitable $C^*$
algebras are made. These are available for all the usual scalar, spinor, 
gauge fields. The kinematical Hilbert spaces constructed here differ from 
the usual Fock spaces. The crucial point of these constructions is to have 
no dependence on background space-time geometry.\cite{std-model} 

\section{Quantum Symmetry Reduction}

One would now like to specialize this frame work to the cosmological
context of highly symmetric space-times. Since the configuration space
variable are now connections, the notion of symmetry requires that
the connection transformed under a symmetry diffeomorphism of $\Sigma$
to be gauge equivalent to the original connection. The first task is to
characterize such symmetric connections precisely. This has already been
done and can be summarized as (simplifying a bit for brevity):\cite{class,bk}\\

If $S$ is a symmetry group (compact Lie group) and $F$ a (Lie) subgroup
of $S$, then \\

(a) $\Sigma \sim B \times S/F \ , B \sim \Sigma/S $ is the space of
orbits of S-action on $\Sigma$ ; \\

(b) Symmetric connections on $\Sigma$ are completely characterized by a
(reduced) connection on $B$ together with a set of (Higgs) scalars on
$B$, possibly satisfying further constraints. \\

For example, for the Bianchi class A models, $S$ is one of the Bianchi
groups with structure constants satisfying $C^I_{JI} = 0$ while $F =
\{e\}$. $\Sigma$ is the group manifold of $S$ while $B$ is a single point
of $\Sigma$. In terms of the Maurer-Cartan 1-forms $\omega^I$ and the
corresponding left invariant vector fields $X_J$ one has,
\begin{equation}
A^i_a = \Phi^i_I \omega^I_a  \ \ \ \ E^a_i = P^I_i X^a_I 
\ \ \ \ \ \ \{\Phi^i_I, P^J_j\} = 8 \pi G \gamma \delta^i_j \delta^J_I
\end{equation}

$\Phi_I^i$ are the scalars, $P_i^I$ are the conjugate momenta, 
$\gamma$ is the Barbero-Immirzi parameter and the indices $i$ ($su(2)$), 
$I$ (Lie algebra of Bianchi group) both take three values. 
If in addition to spatial homogeneity one 
also has isotropy then the scalars satisfy further conditions whose solution 
is $\Phi_I^i = c \delta_I^i$ while the conjugate momenta satisfy $P_i^I = p
\delta_i^I$. The constraint expressions can be likewise simplified and
expressed in terms of the scalars and their conjugates. This is the
classical symmetry reduction. If one proceeds with quantization in a
traditional manner then this is very similar to the usual minisuperspace
quantization apart from the phase space variables being different. The
results are also similar.\cite{kodama} \\

However an alternative quantization is possible. Instead of requiring
the scalars to be well defined operators one can take their
exponentials, the so called point holonomies, as well defined operators.
Then, similar to the general framework of quantum geometry (polymer
representation), one constructs a new $H_{kin}^B$. One can immediately
ask why one should do this? Does this Hilbert have anything to do with
the $H_{kin}^{\Sigma}$ of the full theory? The answer turns out to be
yes! Bojowald and Kastrup show that the (cylindrical) states in $H_{kin}^B$
can be identified with those distributions in $\Omega^*_{\Sigma}$ whose
support consists of precisely the classical (smooth) symmetric 
connections.\cite{bk}
Recall that the physical states of the full theory are supposed to
reside in $\Omega^*_{\Sigma}$. A natural identification of symmetric
(distributional) states of the full theory is via the properties of their 
support. The result shows that such quantum states can be alternatively
be dealt with by working with the cylindrical states of $H_{kin}^B$.
This provides a justification for the alternative (holonomy based) 
quantization of the symmetry reduced theory. It also makes available the
tools of the general framework in a simplified context. This is 
what is referred to as `quantum symmetry reduction'. Note that this is
quite general and not restricted to a cosmological context.

\section{Bianchi Class A models}

The strategy now is step by step adaptation of the general framework.
While majority of steps are identical, there are also crucial new inputs 
needed particularly when additional symmetry such as isotropy is at work.
I am including only the minimal details necessary to communicate the
final results. \\

For classical configurations, point holonomies are just the group
elements obtained by exponentiating the Lie algebra valued scalars,
$u(\Phi_I) \equiv exp\{ \Phi_I^i \tau_i\} \in SU(2)$. Distributional
scalars do just that, they associate with each vertex, an $SU(2)$ group 
element. For general anisotropic case the kinematical Hilbert space turns 
out to be $H_{kin} = L^2( SU(2)^3, d\mu_{Haar}^3 )$.\cite{b1} There are 
three copies since $I$ takes three values. For gauge invariant functions, 
a convenient basis is provided by the usual spin-network functions 
constructed from graphs $\gamma$ in the group manifold of $S$ with a 
single vertex of order 6 and three (closed) edges corresponding to the 
three left invariant vector fields.
Using these one defines the operators corresponding to the conjugate
momenta and then proceeds to build the constraint operators. Following
the Thiemann approach, the Hamiltonian constraint is expressed in terms
of the volume operator together with various commutators of holonomies
with the volume operator. So the main problem is to define a volume
operator and obtain its spectrum. The full spectrum is not available in
the general case. However, for the homogeneous and isotropic models
(Bianchi I and IX), spectrum of volume operator has been determined. \\

When isotropic case is considered, the scalars have to satisfy further
conditions. Naively one would expect that since we now have a single
scalar, spin-network functions associated with graphs with a single closed
edge
(and a single vertex) should suffice. This turns out to be false.
Although the $H_{kin}$ in this case does turn out to be
$L^2(SU(2),d\mu_{Haar})$, it contains more gauge invariant functions
than what a graph with single edge could supply. Bojowald determines the
extra functions needed thereby obtaining an explicit orthonormal basis for
the kinematical Hilbert space. A volume operator is now defined
explicitly and is spectrum obtained.\cite{b2} The eigenfunctions and eigenvalues
are given by (in the gauge invariant sector), 
\begin{eqnarray}
\chi_j(c) & = & \frac{sin((j + \frac{1}{2})c)}{sin(c/2)} \ \ \ \ \  j =
0, 1/2, 1, ...  \nonumber \\
\zeta_j(c) & = & \frac{cos((j + \frac{1}{2})c)}{sin(c/2)} \  \ \ \ \ j =
0, 1/2, 1, ...  \nonumber \\
\zeta_{-\frac{1}{2}} & = & \frac{1}{\sqrt{2} sin( c/2 )} \nonumber \\
V_j & = & (\gamma \ell_p^2)^{3/2} \sqrt{\frac{j (j + \frac{1}{2}) (j + 1)}{27}}
\end{eqnarray}

It turns out to be more convenient to choose a slightly different
orthonormal basis which we now denote as:\cite{absense}
\begin{equation}
|n\rangle \equiv \frac{e^{i \frac{n}{2} c}}{\sqrt{2} sin(c/2)}, \ \ \ \ \ n \in 
{\bf Z}
\end{equation}

These are also eigenstates of the volume operator with eigenvalues
$V_{\frac{|n| - 1}{2}}$. Now the Hamiltonian constraint can be defined by 
its action on these states. In general it has the form: 
\begin{equation}
\hat{H} | n \rangle = \sum_{k = -L}^{L} A_{n - k}^k | n - k \rangle .
\end{equation}

The physical states, $| s \rangle = \sum_n s_n |n\rangle $, which belong to the
kernel of $\hat{H}$, have conditions on the coefficients $s_n$ of the
form, 
\begin{equation}
\sum_{k = -L}^{L} A_n^k s_{n + k} = 0 .
\end{equation}

This is a difference equation for the $s_n$ coefficients. The order of
the equation, $2 L$, depends on the Bianchi type while the coefficients
$A_n^k$ depend on the details of the Hamiltonian constraint and the
factor ordering chosen.\cite{b3,b4,isotropic} For the flat isotropic model, the order 
of the equation is 16. The ({\it non-}symmetric) ordering chosen is such 
that the coefficient of $s_0$ always vanishes. One can introduce matter 
sector and add its contribution to the Hamiltonian constraint. Denoting matter
symbolically by $\varphi$ the equations are modified by putting in a
$\varphi$ dependence in $s_n$ and adding a term of the form
$\hat{H}_{\varphi}(n) s_n (\varphi)$. At this stage, particular form of
matter and its couplings are not detailed. It is sufficient to note that the
matter part of the Hamiltonian operator is diagonal with respect to the
(gravitational) states $|n\rangle$. This specifies the physical states of 
the loop quantum cosmology.\cite{absense,unique}

\section{Dynamical Interpretation, Absence of Singularity and Uniqueness
of Solution}

In order to interpret the physical states, in particular to explore
what `happens' to the classical singularity, one needs to view the
physical states obtained above as solutions of a `time evolution' equation.
In a generally covariant theory of space-time there is no external,
inert `time' variable. One can at best single out one of the degrees of
freedoms as a `clock' and view functions of this (gravitational or matter) 
degree of freedom as functions of `time'. While there is no a priori given 
clock, in a given context there can be a natural choice. For instance,
for the FRW cosmologies, the scale factor in the metric is one such choice. 
It is in terms of this choice that one says that there is a singularity
at the vanishing value of the scale factor (curvatures blow up). In the
present formulation, absolute value of the conjugate momentum, $p$,
corresponds to the square of scale factor. The spectrum of $|\hat{p}|$ 
consists of eigenvalues $(j + 1/2)$ with eigenfunctions $\chi_j(c), \zeta_j(c)$.
The zero eigenvalue of $|\hat{p}|$ occurs for $j = -1/2$ which is a 
non-degenerate eigenvalue. (The volume operator consists of this operator 
together with another commuting operator with eigenvalues $j(j + 1)$.
This makes the zero eigenvalue of the volume operator three fold
degenerate.) In terms of the $|n\rangle$ basis, $n = 0$ corresponds to
vanishing scale factor. One therefore takes, the label $n \in
{\bf Z}$ as a time label. The physical state condition can now be viewed 
as specifying a discrete time evolution.\cite{b4,absense,unique}\\

Restricting to the isotropic, flat models, the evolution equation is of
order 16 (and consists of the $n \pm 8, n \pm 4$ and $n$ terms only). The
coefficients are such that $A_n^k = 0$ if and only if $n + k = 0$. Furthermore,
$\hat{H}_{\varphi}(n = 0) = 0$ as well. This has two consequences. The
$s_0(\varphi)$ never appears in the equation. Therefore it is neither
determined by nor determines any other $s_n(\varphi)$. The state
$|0\rangle$ is thus orthogonal to all other physical states (evolving
solutions). This is precisely the state which corresponds to the zero
eigenvalue of the scale factor operator. Thus one sees that unlike the
classical case where evolution through vanishing scale factor is not
possible, the quantum evolution equation does not suffer such a
breakdown. In this sense the classical singularity is absent in the evolving 
states. The second consequence is that one gets a conditions on the initial 
data - the choice of 16 coefficients $s_{-16}, s_{-15}, ... s_{-1}$ (say). 
This is because the equation can not determine $s_0$ since its coefficient
vanishes. Thus one can conclude that the classical singularity is
avoided by the evolving solutions and that there are 15 instead of 16
independent solutions.\cite{absense,unique} \\

This seems worse than the standard quantum cosmology with only two
independent solutions! Quite independently, one has also to understand
the discrete time evolution in terms of more familiar continuous time
evolution, at least when one hopes to be in classical regime. Clearly,
the volume and the scale factor eigenvalues become large compared to the
Planck scale values when $n \gg 1$. One expects a solution to be
classically interpretable when for large $n$, $s_{n + m}(\varphi)$ can
vary significantly from $s_n(\varphi)$ when $m$ is also large
compared to 1 but remains almost the same when $m$ is comparable to 1.
One can now ask: how many solutions exhibit this behavior? In a
precise formulation of such a behavior, the Barbero-Immirzi parameter
$\gamma$ come very handy. Note that for large $n = 2j + 1$, $V_j \to
(\gamma \ell_p^2 |n|/6 )^{3/2} \sim (a^2)^{3/2}$. We can use this to define
the scale factor as a function of $n$ as: $a^2(n) \equiv (n \gamma)
\ell_p^2 /6 $. Now change in $a^2(n)$ as $n$ changes by $1$ will be
infinitesimal if $\gamma \ell_p^2 \to 0$. Thus we can mimic continuous
evolution with respect to the scale factor by considering the formal limit 
$\gamma \to 0 , n \to \infty$ keeping $\gamma n$ a constant. We are
keeping $\ell_p$ fixed in this and so are in the quantum domain still.
If a solution $\{s_n(\varphi)\}$ has a limiting value in the above limit
then it is said to be `pre-classical'.\cite{unique} The question now 
becomes as to how many evolving solutions are pre-classical? For large $n$, the
equations itself becomes an equation with constant coefficients and is easy 
to analyze. The result is that, of the sixteen solutions, only two are
pre-classical. There is still the constraint on the initial conditions
which reduces these two to a single solution. Thus while there are many
evolving solutions (all avoiding the singularity), only one of these can
mimic a classical evolution.\cite{unique} \\

From the black hole entropy computations, $\gamma$ is fixed to be a
constant of order one. The $\gamma \to 0$ limit noted above is to be
thought of as a formal device to single out solutions mimicking
classical evolution. The presence of the Barbero-Immirzi parameter
however offers a new limit to be explored apart from the classical limit
with $\ell_p \to 0$. Bojowald has further shown that in the $\gamma \to
0$ limit, referred to as `continuum limit' in the sense that the discrete
structure of quantum geometry can be ignored, one recovers the standard
quantum cosmology.\cite{semiclassical} This shows also that it is the 
specific, discrete structure of quantum geometry that is responsible for 
avoiding the singularity and selecting a unique (pre-classical) solution and 
not just any quantization procedure.

\section{Concluding Remarks}
This is a first explicit application of the highly abstract
framework of quantum geometry to a context of physical interest,
particularly addressing the issue of classically indicated singularity.
Not only does it meet the expectation that in a quantum theory of
gravity, classically
indicated singularities should be absent, it provides a quantitative
means to estimate how rapidly quantum geometry picture goes over to the
classical picture. In the quantization procedure, many conceptual and
technical assumptions have gone in (eg choice of polymer representation).
In a sense the classical space-time picture has been `mutilated' quite a
bit. Therefore it could have happened that the none of final set of solutions
could be interpreted in classical terms thereby recovering the
classical theory. Not only this does {\it {not}} happen but one gets a unique
solution displaying the classical picture. It also brings out an
intriguing role played by the Barbero-Immirzi parameter. \\

{\underline{Acknowledgments:}}
It is a pleasure to thank the organizers for the invitation and IUCAA
for the excellent atmosphere and hospitality. I would like to thank
Martin Bojowald for his useful comments on an earlier draft.

\end{document}